\documentclass[aps,preprint,onecolumn,noshowpacs,nofootinbib]{revtex4}
\usepackage{amssymb}
\usepackage{amsmath}
\usepackage{amsfonts}
\usepackage{mathrsfs}
\usepackage{color}
\usepackage{ulem}
\usepackage{physics}
\usepackage[english]{babel}

\usepackage{graphicx}

\graphicspath{{figures/}}

\begin{document}

\title{Formal equivalence between Maxwell equations and the de Broglie-Bohm theory for two-dimensional optical microcavities }
\author{Aur\'elien Drezet$^{1}$,Bernard Michael Nabet$^{2}$}
%\institute{                    
\affiliation{$^1$Univ. Grenoble Alpes, CNRS, Grenoble INP, Institut Neel, F-38000 Grenoble, France}
\email{aurelien.drezet@neel.cnrs.fr}
\affiliation{$^2$Technion–Israel Institute of Technology, Humanities and Arts Department, Haifa, Israel}

\begin{abstract}
We analyze the formal equivalence between the electromagnetic energy conservation law derived from Maxwell's equations in an optical microcavity and the conservation of a probability fluid associated with the de Broglie-Bohm theory for an effective massive particle describing a photon in this cavity. This work is part of a critical analysis of recent experiments [Nature \textbf{643}, 67-72 (2025)] carried out with a view to refuting the de Broglie-Bohm theory. Furthermore, the consequences of our analysis for microphotonics go far beyond these experiments. In particular, extensions that take into account photon spin and stochastic aspects associated with radiative or absorption losses are considered.
\end{abstract}

\maketitle

The de Broglie-Bohm pilot wave theory, i.e., Bohmian mechanics \cite{Valentini,Bohm1952,Hiley,HollandB}, is originally an alternative interpretation of quantum mechanics that preserves the notion of classical determinism and the notion of trajectories for particles.\\
\indent It can be rigorously demonstrated that, given certain statistical assumptions about the so-called quantum equilibrium regime, Bohmian theory is capable of reproducing the entirety of quantum mechanics, at least in the non-relativistic regime. In particular, probabilistic predictions and quantum measurement theory can be encompassed and fully justified within a Bohmian framework \cite{HollandB,Durr,Teufel}.\\  
\indent However, there is no consensus regarding relativistic quantum theories, including quantum field theories (QFT) or even quantum optics. Among the various approaches that have been attempted in QFT, we mention Bohm's initial work based on a functional representation of the quantum electromagnetic field coupled to a (non-relativistic) description of fermions treated as particles \cite{Bohm1952b,Hiley,Struyve2010}. We also mention attempts to develop a Bohmian quantum electrodynamics (QED) for the Dirac equation based on the old ideas of an infinite sea of negative-energy particles \cite{Hiley,Colin,Deckert}. Finally, we mention works inspired by Bell's research, which introduces stochastic elements into the Bohmian approach in order to take into account the creation and annihilation of particles in QFT \cite{Bell,Sudbery,Vink,Tumulka}.\\
\indent Beyond these fundamental and interpretative questions, there is no doubt that Bohmian formalism, based on a hydrodynamic representation known as Madelung's representation of wave equations (Schr\"odinger, Klein-Gordon, Dirac), is an interesting tool for describing wave propagation phenomena in complex environments. In optics, and particularly in nano/microphotonics, a Bohmian analogy of wave propagation is certainly possible, even in the classical regime based on pure Maxwell equations. In such a theory, the speed of the Bohmian fluid must intuitively be related to the propagation of the energy fluid deduced from Maxwell's equations via Poynting's theorem. This is what we will show below.\\
\indent More precisely, in this paper, we propose such a Bohmian hydrodynamic formulation for a classical electromagnetic field propagating in a planar optical microcavity. This work is strongly motivated by recent results published by Sharoglazova et al. \cite{Klaers2025,Klaers2023} concerning the Bohmian  velocity in a microcavity. Based on an experiment involving structured waveguides in a planar optical microcavity, these authors sought to demonstrate an inconsistency in Bohmian theory in the evanescent wave regime using an optical analog of the tunnel effect. We recently proposed a clear refutation of this claim \cite{Drezet2025} based principally on the existence of radiative leaks neglected in the authors' analysis, which allows us to unambiguously determine a Bohmian velocity even in their experiment [In \cite{Drezet2025}, we highlighted a second effect linked to the finite duration of the excitation pulses, which in itself is sufficient to invalidate the results of \cite{Klaers2025} and save Bohmian mechanics from refutation. However, as the optical response is dominated by radiative leakage, we will not consider it further here].  However, underlying this work is the assumption that a coherent Bohmian formalism exists in such a microcavity.
In the model published to date, we start from a formal analogy between the wave equations in the microcavity and a particle with effective mass $m$ due to the particular paraxial regime used. This strongly motivates an equivalence between the wave equations of the electromagnetic field and the non-relativistic Schrödinger equation for a massive particle. However, the Bohmian theory assumed in these works is not exactly deduced from Maxwell's equations, and the link between energy propagation and Bohmian propagation is not firmly established. We emphasize that this in no way invalidates previous analyses made to account for observations, as the analysis based on the paraxial optical regime is sufficiently well established at such a phenomenological level. In other words, readers wishing to find a refutation of claim \cite{Klaers2025} can simply read the self-sufficient work \cite{Drezet2025}. However, in a second step, this strongly suggests a more detailed theoretical analysis of this type of paraxial analogy and the equivalence between Bohmian dynamics and energy transport in a 2D microcavity.  Furthermore, given the importance of the problem more generally for the study of optical microcavities, we propose here to justify the existence of such a Bohmian model for a 2D optical analogy. 

\indent In this context, it is essential to note that de Broglie, guided by Einstein's work, had suggested as early as 1924-25  \cite{deBroglie1924,deBroglie1925} (i.e., before the discovery of Schrödinger's equation) that the speed of photons is given by the ratio between the Poynting vector $\mathbf{S}=\mathbf{E}\times \mathbf{B}$, which defines the local energy flux, and the local energy density $u=\frac{\mathbf{E}^2+\mathbf{B}^2}{2}$:        
\begin{eqnarray}
\mathbf{v}(\mathbf{x},t)=\frac{\mathbf{S}(\mathbf{x},t)}{u(\mathbf{x},t)}=2\frac{\mathbf{E}(\mathbf{x},t)\times \mathbf{B}(\mathbf{x},t)}{\mathbf{E}^2(\mathbf{x},t)+\mathbf{B}^2(\mathbf{x},t)}
\end{eqnarray} where $\mathbf{E}(\mathbf{x},t)$ and $\mathbf{B}(\mathbf{x},t)$ are respectively the electric and magnetic field vector at point $\mathbf{x}$ and time $t$ obeying Maxwell equations. Such ideas were also introduced by Slater  \cite{Slater}.\\
\indent The motivation for this formula is, of course, the local law of conservation of energy $\partial_t u=-\boldsymbol{\nabla}\cdot\mathbf{S}$. Similarly, in the case of the Schr\"odinger equation for a single particle of mass $m$, according to Bohmian mechanics, we can define a local velocity for the particle using the ratio between the probability current $\mathbf{J}(\mathbf{x},t)=\frac{1}{m}\textrm{Im}[\Psi^\ast(\mathbf{x},t)\boldsymbol{\nabla}\Psi(\mathbf{x},t)]$ and the local probability density $\rho(\mathbf{x},t)=\Psi^\ast(\mathbf{x},t)\Psi(\mathbf{x},t)$, which gives the de Broglie-Bohm guidance formula:   
\begin{eqnarray}
\mathbf{v}_{dBB}(\mathbf{x},t)=\frac{\mathbf{J}(\mathbf{x},t)}{\rho(\mathbf{x},t)}=\frac{1}{m}\textrm{Im}[\frac{\boldsymbol{\nabla}\Psi(\mathbf{x},t)}{\Psi(\mathbf{x},t)}]
\end{eqnarray} where  $\Psi(\mathbf{x},t)$ is the wavefunction obeing the nonrelativistic Schr\"odinger equation. \\
\indent  Here again, an obvious motivation for this formula lies in the law of local conservation $\partial_t \rho=-\boldsymbol{\nabla}\cdot\mathbf{J}$. However, it is important to note some fundamental differences between Maxwell's case and Schr\"odinger's: \\
\indent i) First of all, in the case of Schr\"odinger's equation, we are dealing with the current and density of a probability fluid, whereas in Maxwell's case, we are dealing with an energy fluid.  This is central from a physical point of view. Moreover, from a relativistic point of view, the energy fluid does not define a current four-vector but is part of an energy-momentum tensor $T^{\mu\nu}$. The velocity obtained therefore cannot have fundamental relativistic validity. In contrast, in the case of Schr\"odinger's equation, the guiding formula can easily be generalized to the Dirac or Klein Gordon equation, where the current and density form a current four-vector $J^\mu:=[\rho,\mathbf{J}]$ with the correct Lorentz covariant meaning (note however that for the Klein-Gordon case the current $J^\mu$ is not always time-like \cite{HollandB}).\\
\indent In this work, where we are only interested in hydrodynamic optical analogies for an optical microcavity, we will not seek to preserve the relativistic covariance of the problem, and we can ignore this issue.\\
\indent ii) Another fundamental difference is, of course, that Maxwell's theory is based on real electric and magnetic fields, whereas Schrödinger's wave function is a complex field.  In the regime under consideration, we will show that the introduction of a complex electromagnetic field is naturally required if we assume a light source that randomly emits pulses of identical nature and form but shifted in time by a statistically distributed variable delay. Furthermore, the transition from a field with multiple E and B components to a single wave function will also be justified by the symmetries and approximations of the problem under consideration.\\
\indent iii) A final important difference between Maxwell's equations and Schr\"odinger's equation concerns linearity and the absence of a source term in the latter. In particular, the conservation law for electromagnetic energy generally contains a source term associated with the electric current interacting with the electric field: $\partial_t u=-\boldsymbol{\nabla}\cdot\mathbf{S}- \mathbf{J}_e\cdot\mathbf{E}$. Such a source term is absent from the probabilistic conservation law in the Schr\"odinger equation because the number of particles is conserved in the non-relativistic regime. However, this difference disappears in the relativistic regime, particularly in QED, and Bohmian models that take this fact into account have been developed. We will not have to consider this problem in full generality because we will limit ourselves to linear optical media where electric currents are linked to local polarization. Note however that including loss and absorption by the optical medium requires such a stochastic approach as discussed below.\\
\indent  The structure of the paper is as follows: In Section \ref{sec2}, we provide a detailed formulation of Maxwell’s equations in terms of complex electromagnetic fields. In particular, we introduce the Poynting vector and the energy density associated with this complex field, and we discuss the issue of absorption. In Section \ref{sec3}, we justify the emergence of an effective Schr\"odinger equation for the Maxwell field in an ideal planar microcavity. These results are then extended to the case of a lossy cavity in Section \ref{sec4}. In both Sections \ref{sec3} and \ref{sec4}, we develop a Bohmian analogy and demonstrate that the in-plane velocity of the energy fluid within the cavity corresponds precisely to the Bohmian velocity derived from the Schr\"odinger equation. Section \ref{sec4} also discusses possible extensions of the model including stochasticity or spin. Finally, in Section \ref{sec5}, we revisit the discussion surrounding the claims made in \cite{Klaers2025}, in light of our previous comment \cite{Drezet2025} and the present work. We conclude by summarizing the implications of our findings for integrated optics and Bohmian mechanics.
%%%%%%%
\section{The Poynting theorem for  complex effective electromagnetic fields}\label{sec2}
We start from Maxwell's equations in the presence of electric polarization $\mathbf{P}$ and magnetic polarization $\mathbf{M}$:
\begin{eqnarray}
\boldsymbol{\nabla}\times \mathbf{E}=-\partial_t \mathbf{B} &, \boldsymbol{\nabla}\cdot \mathbf{B}=0\nonumber\\
\boldsymbol{\nabla}\times \mathbf{B}=\partial_t \mathbf{E} + \partial_t \mathbf{P} +\boldsymbol{\nabla}\times \mathbf{M} &, \boldsymbol{\nabla}\cdot \mathbf{E}=-\boldsymbol{\nabla}\cdot \mathbf{P}\label{max1}
\end{eqnarray}
In a linear regime, we assume the following in order to satisfy causal constraints: 
\begin{eqnarray}
\mathbf{P}(\mathbf{x},t)=\int_0^{+\infty}d\tau \chi_E(\mathbf{x},\tau)\mathbf{E}(\mathbf{x},t-\tau)\nonumber\\
\mathbf{M}(\mathbf{x},t)=\int_0^{+\infty}d\tau \chi_M(\mathbf{x},\tau)\mathbf{B}(\mathbf{x},t-\tau)\label{pola}
\end{eqnarray}
with $\chi_{E,M}(\mathbf{x},\tau)$ linear susceptibilities depending on a time delay $\tau$ associated with dispersion and dissipation (note that for simplicity, we assume here spatially local media).\\
\indent At this point, we introduce statistical assumptions and assume that the electromagnetic state is described by a linear superposition of pulses such that for each pulse we use a Fourier decomposition of the type
\begin{eqnarray}
\mathbf{E}_\delta(\mathbf{x},t)=\int_0^{+\infty}d\omega \mathbf{E}_\omega(\mathbf{x})e^{-i(\omega t+\delta(\omega))} +cc.
\end{eqnarray} and similarly for the magnetic field.
 Basically, we have in fact introduced a phase delay $\delta(\omega)$ which is assumed to vary randomly from one pulse to another. For greater precision, we limit ourselves to the hypothesis $\delta(\omega)=\delta_0$ so that we can write  
\begin{eqnarray}
\mathbf{E}_\delta(\mathbf{x},t)=\boldsymbol{\mathcal{E}}(\mathbf{x},t)e^{i\delta_0} +cc.
\end{eqnarray}  with  $\boldsymbol{\mathcal{E}}(\mathbf{x},t)=\int_0^{+\infty}d\omega \mathbf{E}_\omega(\mathbf{x})e^{-i\omega t}$ a complex electric field (and similarly for the magnetic field).\\
\indent Using this assumption about random phases, it is easy to convince oneself that each of Maxwell's equations must be sastified for complex fields $\boldsymbol{\mathcal{E}}$, $\boldsymbol{\mathcal{B}}$, $\boldsymbol{\mathcal{P}}$ and $\boldsymbol{\mathcal{M}}$ i.e.:   
   \begin{eqnarray}
\boldsymbol{\nabla}\times \boldsymbol{\mathcal{E}}=-\partial_t \boldsymbol{\mathcal{B}} &, \boldsymbol{\nabla}\cdot \boldsymbol{\mathcal{B}}=0\nonumber\\
\boldsymbol{\nabla}\times \boldsymbol{\mathcal{B}}=\partial_t \boldsymbol{\mathcal{E}} + \partial_t \boldsymbol{\mathcal{P}} +\boldsymbol{\nabla}\times \boldsymbol{\mathcal{M}} &, \boldsymbol{\nabla}\cdot \boldsymbol{\mathcal{E}}=-\boldsymbol{\nabla}\cdot \boldsymbol{\mathcal{P}}\label{max2}
\end{eqnarray}  and the linear equations Eq.~\ref{pola} are left unchanged with real fields replaced by complex fields.\\                  
\indent In this new framework, we recall that according to Poynting's theorem derived from Eq.~\ref{max1}, we have
\begin{eqnarray}
\mathbf{E}\partial_t\mathbf{D}+\mathbf{H}\partial_t\mathbf{B}=-\boldsymbol{\nabla}\cdot[\mathbf{E}\times\mathbf{H}] \label{po1}
\end{eqnarray}
where we have introduced the fields $\mathbf{D}=\mathbf{E}+\mathbf{P}$, $\mathbf{H}=\mathbf{B}-\mathbf{M}$. 
In this formula, the Poynting vector associated with the local energy flux is defined by $\mathbf{S}=\mathbf{E}\times\mathbf{H}$. We can easily extend Poynting's theorem to the complex Maxwell field directly from Eq.~\ref{max2}
\begin{eqnarray}
2\textrm{Re}[\boldsymbol{\mathcal{E}}^\ast\partial_t\boldsymbol{\mathcal{D}}+\boldsymbol{\mathcal{H}}^\ast\partial_t\boldsymbol{\mathcal{B}}]=-\boldsymbol{\nabla}\cdot2\textrm{Re}[\boldsymbol{\mathcal{E}}^\ast\times\boldsymbol{\mathcal{H}}]\label{po2}
\end{eqnarray}
Alternatively, we can obtain this formula by introducing a probability distribution  $P(\delta)$ for the random phase $\delta:=\delta_0$ and defining the mean  $\langle A_\delta(\mathbf{x},t)\rangle=\int d\delta P(\delta)A_\delta(\mathbf{x},t) $. We thus get  by averaging Eq.~\ref{po1} $\langle \mathbf{E}_\delta\partial_t\mathbf{D}_\delta+\mathbf{H}_\delta\partial_t\mathbf{B}_\delta\rangle=2\textrm{Re}[\boldsymbol{\mathcal{E}}^\ast\partial_t\boldsymbol{\mathcal{D}}+\boldsymbol{\mathcal{H}}^\ast\partial_t\boldsymbol{\mathcal{B}}]$  and similarly $\langle \mathbf{S}_\delta(\mathbf{x},t)\rangle=2\textrm{Re}[\boldsymbol{\mathcal{E}}^\ast\times\boldsymbol{\mathcal{H}}]:=\boldsymbol{\mathcal{S}}$ which is leading once again to Eq.~\ref{po2}.\\
\indent In order to be more specific about the type of model considered here, we introduce the electric permittivity $\varepsilon_\omega$ and the magnetic permeability $\mu_\omega$ defined by relations 
\begin{eqnarray}
\varepsilon_\omega(\mathbf{x})=1+\int_0^{+\infty}d\tau \chi_E(\mathbf{x},\tau)e^{i\omega\tau}\nonumber\\
\mu_\omega(\mathbf{x})=\frac{1}{1-\int_0^{+\infty}d\tau \chi_M(\mathbf{x},\tau)e^{i\omega\tau}}
\end{eqnarray}
and following Landau and Lifschitz \cite{Landau}, we work in the weakly dispersive regime around a mean frequency $\omega_0$.  According to these authors, writing $\boldsymbol{\mathcal{E}}(\mathbf{x},t)=\tilde{\boldsymbol{\mathcal{E}}}(\mathbf{x},t)e^{-i\omega_0 t}$ where the envelope field $\tilde{\boldsymbol{\mathcal{E}}}$ varies slowly compared to $e^{-i\omega_0 t}$, we deduce at the first order approximation: 
\begin{eqnarray}
\boldsymbol{\mathcal{E}}^\ast\partial_t\boldsymbol{\mathcal{D}}\simeq -i\omega_0\varepsilon_{\omega_0}|\boldsymbol{\mathcal{E}}|^2 +\frac{d(\omega\varepsilon_\omega)}{d\omega}|_{\omega_0}\tilde{\boldsymbol{\mathcal{E}}}^\ast\partial_t\tilde{\boldsymbol{\mathcal{E}}}
\end{eqnarray}
which allows us to rewrite Poynting's theorem Eq.~\ref{po2} in the approximate form:          
\begin{eqnarray}
\frac{d(\omega\varepsilon'_\omega)}{d\omega}|_{\omega_0}\partial_t|\boldsymbol{\mathcal{E}}|^2+\frac{d(\omega\mu'_\omega)}{d\omega}|_{\omega_0}\partial_t|\boldsymbol{\mathcal{H}}|^2\nonumber\\\simeq-\boldsymbol{\nabla}\cdot2\textrm{Re}[\boldsymbol{\mathcal{E}}^\ast\times\boldsymbol{\mathcal{H}}]-2\omega_0\varepsilon''_{\omega_0}|\boldsymbol{\mathcal{E}}|^2-2\omega_0\mu''_{\omega_0}|\boldsymbol{\mathcal{H}}|^2\nonumber\\-2\textrm{Im}[\tilde{\boldsymbol{\mathcal{E}}}^\ast\partial_t\tilde{\boldsymbol{\mathcal{E}}}]\frac{d(\omega\varepsilon''_\omega)}{d\omega}|_{\omega_0}-2\textrm{Im}[\tilde{\boldsymbol{\mathcal{H}}}^\ast\partial_t\tilde{\boldsymbol{\mathcal{H}}}]\frac{d(\omega\mu''_\omega)}{d\omega}|_{\omega_0}\label{po3}
\end{eqnarray} where $\varepsilon'_\omega=\textrm{Re}[\varepsilon_\omega]$,$\varepsilon''_\omega=\textrm{Im}[\varepsilon_\omega]$ and similarly for $\mu_\omega$.\\
\indent This equation can be interpreted physically. On the left-hand side appears the time derivative of the well-known electromagnetic energy density $u=\frac{d(\omega\varepsilon'_\omega)}{d\omega}|_{\omega_0}|\boldsymbol{\mathcal{E}}|^2+\frac{d(\omega\mu'_\omega)}{d\omega}|_{\omega_0}|\boldsymbol{\mathcal{H}}|^2$. On the right-hand side, apart from the usual divergence term involving the Poynting vector $\boldsymbol{\mathcal{S}}$, there are terms related to the dissipation of the polarized medium $-2\omega_0\varepsilon''_{\omega_0}|\boldsymbol{\mathcal{E}}|^2-2\omega_0\mu''_{\omega_0}|\boldsymbol{\mathcal{H}}|^2$. Since the imaginary parts $\varepsilon''_\omega,\mu''_{\omega}$ are always positive by virtue of the Kramers-Kronig relations for a dissipative system, this corresponds to absorption by the polarizable medium. The last terms on the right-hand side are more difficult to interpret and are considered negligible by Landau and Lifschitz \cite{Landau}.\\ 
\indent We point out that our derivation of Poynting's theorem for a polarizable medium is more general than those usually given in the literature, because we use ensemble averages in the delay space $\delta$ rather than time averages. This allows us to introduce complex electromagnetic fields and retain time dependencies in our expressions.\\
\indent To conclude this first part, we have derived Poynting's theorem 
\begin{eqnarray}
\partial_t u=-\boldsymbol{\nabla}\cdot\boldsymbol{\mathcal{S}}-\mathcal{W}\label{conse}
\end{eqnarray}
with \begin{eqnarray}
u=\frac{d(\omega\varepsilon'_\omega)}{d\omega}|_{\omega_0}|\boldsymbol{\mathcal{E}}|^2+\frac{d(\omega\mu'_\omega)}{d\omega}|_{\omega_0}|\boldsymbol{\mathcal{H}}|^2\nonumber\\
\boldsymbol{\mathcal{S}}=2\textrm{Re}[\boldsymbol{\mathcal{E}}^\ast\times\boldsymbol{\mathcal{H}}]
\end{eqnarray} and a dissipation term $\mathcal{W}=2\omega_0\varepsilon''_{\omega_0}|\boldsymbol{\mathcal{E}}|^2+2\omega_0\mu''_{\omega_0}|\boldsymbol{\mathcal{H}}|^2$. This allows us to introduce a velocity for the energy flow with the definition:
\begin{eqnarray}
\mathbf{v}(\mathbf{x},t)=\frac{\boldsymbol{\mathcal{S}}(\mathbf{x},t)}{u(\mathbf{x},t)}=\frac{2\textrm{Re}[\boldsymbol{\mathcal{E}}^\ast\times\boldsymbol{\mathcal{H}}]}{\frac{d(\omega\varepsilon'_\omega)}{d\omega}|_{\omega_0}|\boldsymbol{\mathcal{E}}|^2+\frac{d(\omega\mu'_\omega)}{d\omega}|_{\omega_0}|\boldsymbol{\mathcal{H}}|^2}\label{dBB1}
\end{eqnarray}
%%%%%%%%%%%%%%%%%%%%%%%%
\section{Bohmian formalism for an ideal two-dimensional optical microcavity}\label{sec3}
\indent Having developed a hydrodynamic representation for the energy fluid based on a complex electromagnetic field, this suggests, by direct analogy with what has been done for the Schr\"odinger or Dirac equation, defining a velocity for a light particle using Eq.~\ref{dBB1}.   However, at this stage, we note that the presence of a loss term $\mathcal{W}$ is at odds with a purely deterministic approach. Indeed, the presence of $\mathcal{W}(\mathbf{x},t)$ implies that the trajectories of light particles can in some cases be interrupted at space-time point $x^\mu:=[t,\mathbf{x}]$ due to absorption by the dissipative medium.\\
\indent Moreover, it is possible to develop an approach similar to that of Bell and taken up by Tumulka et al. to produce a stochastic Bohmian version of QFT \cite{Bell,Sudbery,Vink,Tumulka} (this will be briefly discussed in the next section). In the present section we will neglect losses and absorption and suppose a purely  deterministic hydrodynamic (Bohmian) picture.\\
\indent In this paper, we consider a simplified model of a planar optical cavity used in \cite{Klaers2025}. First, we neglect dispersion and dissipation and consider only a dielectric medium with constant permittivity $\varepsilon$ located between two parallel perfect plane mirrors separated by a distance $D_0$ along the $z$ axis. Moreover, one of the mirrors is weakly structured, and we consider that the distance between the mirrors changes as $D(x,y)=D_0+\delta D(x,y)$ where $|\delta D(x,y)|/D_0 \ll 1$. The wave equations for the electromagnetic fields in the cavity are then written as 
\begin{eqnarray}
[\varepsilon\partial_t^2-\boldsymbol{\nabla}^2]\left( \begin{array}{c} \boldsymbol{\mathcal{E}}\\ \boldsymbol{\mathcal{B}}\end{array}\right)=0.
\end{eqnarray}
Furthermore, we work in the paraxial regime and assume that the electromagnetic field is mainly in the $x-y$ plane parallel to the mirror, while propagation occurs essentially along the $z$ axis. If we work at a fixed frequency $\omega$, this generates a standing wave along the $z$ axis. To simplify we write $\boldsymbol{\mathcal{E}}=\hat{\mathbf{e}}\Phi(x,y,z)e^{-i\omega t}$ with $\hat{\mathbf{e}}$ a  unit constant polarization vector parallel to the $x-y$ plane. The vector is assumed to be real, i.e., $\hat{\mathbf{e}}\in \mathbb{R}^3$. Note that the theory can easily be extended to more complex cases with two wave functions $\Phi_{\pm}$ forming a spinor. We will briefly analyze this possibility at the end of section \ref{sec4}. The wave equation reads $\varepsilon \omega^2\Phi+\partial^2_z \Phi+\boldsymbol{\nabla}_{||}^2\Phi=0$  
(where $\boldsymbol{\nabla}_{||}$ is the gradient operator in the $x-y$ plane) with the approximate solution:
\begin{eqnarray}
\Phi(x,y,z)=\Psi_E(x,y)f_{x,y}(z)
\end{eqnarray}
assuming \begin{eqnarray}
f_{x,y}(z)=\sin{(\frac{q\pi}{D(x,y)}z)}\label{modez}
\end{eqnarray} with  $q\in\mathbb{N}$. The z-component of the wave vector reads $k_z(x,y)=\frac{q\pi}{D(x,y)}\simeq \frac{q\pi}{D_0}-\frac{q\pi\delta D(x,y)}{D_0^2}$. In particular for $\delta D=0$ and $\Psi$ a constant we have the dispersion relation $\varepsilon \omega^2=(\frac{q\pi}{D_0})^2$. This allows us to introduce an effective mass $m=\frac{1}{\sqrt{\varepsilon}}\frac{q\pi}{D_0}$ and we write $k_z(x,y)=\sqrt{\varepsilon}(m+V(x,y))$ where $V(x,y)=-\frac{1}{\sqrt{\varepsilon}}\frac{q\pi\delta D(x,y)}{D_0^2}$ is an effective potential for the in plane motion ($|V|/m\ll 1$). We are considering frequency $\omega$ close to $m$ and write $\omega=m+E$ (i.e., $E/m\ll 1$). The wave equation becomes:     
\begin{eqnarray}
[\varepsilon (m+E)^2+\varepsilon(m+V(x,y))^2+\boldsymbol{\nabla}_{||}^2]\Psi_E(x,y)=0
\end{eqnarray}  and in the `nonrelativistic' approximation $E/m\ll 1$,$|V|/m\ll 1$ this leads to the  time independent effective Schr\"odinger equation:
\begin{eqnarray}
[2\varepsilon m(E-V(x,y))+\boldsymbol{\nabla}_{||}^2]\Psi(x,y)\simeq 0.
\end{eqnarray}
We can easily extend the solution to time-dependent problems by constructing wave packets $\Psi(x,y,t)=\int dE \Psi_E(x,y)e^{-iEt}$ such that $E/m\ll 1$. We then have the time dependent Schr\"odinger equation
\begin{eqnarray}
i\partial_t\Psi(x,y,t)=-\frac{\boldsymbol{\nabla}_{||}^2}{2\varepsilon m}\Psi(x,y,t)+V(x,y)\Psi(x,y,t)
\end{eqnarray}
and we note that including the prefactor $e^{-im t}$ would just add a mass term $m\Psi$ to the right hand side of the equation. We stress that we didn't consider the role of the transversality condition $\boldsymbol{\nabla}\cdot\bold{\mathcal{E}}=0$. We will back to this issue in Section \ref{sec4}.\\
\indent This Schr\"odinger equation clearly motivates a 2D Bohmian analogy with the inplane guidance formula for the fluid velocity:
 \begin{eqnarray}
\mathbf{v}_{dBB,||}(x,y,t)=\frac{1}{\varepsilon m}\textrm{Im}[\frac{\boldsymbol{\nabla}_{||}\Psi(x,y,t)}{\Psi(x,y,t)}]
\end{eqnarray}
The aim of work \cite{Klaers2025} was clearly to criticize the possibility of developing such a Bohmian scheme in a coherent manner so as to explain the data observed in an evanescent field in the $x-y$ plane. However, here we can justify the Bohmian guiding formula using Poynting's theorem, which effectively nullifies the doubts and criticisms claimed in \cite{Klaers2025}.\\
\indent For this purpose we use the previous solution $\Psi(x,y,t)$ and introduce the electric  field $\boldsymbol{\mathcal{E}}(x,yz,t)=\hat{\mathbf{e}}\Psi(x,y,t)e^{-imt}\sin{\chi}$ with $\chi=\sqrt{\varepsilon}(m+V)z$. From  Maxwell equations we have $\boldsymbol{\nabla}\times \boldsymbol{\mathcal{E}}=-\partial_t \boldsymbol{\mathcal{B}} \simeq im \boldsymbol{\mathcal{B}} $  which in turn allows us to define the magnetic field by the equation:  
\begin{eqnarray}
\boldsymbol{\mathcal{B}}\simeq\frac{e^{-imt}}{im}[\sin{\chi}\boldsymbol{\nabla}_{||}\Psi\times\hat{\mathbf{e}}+\sqrt{\varepsilon}z\cos{\chi}\Psi\boldsymbol{\nabla}_{||}V\times\hat{\mathbf{e}}\nonumber\\
+\sqrt{\varepsilon}(m+V)\cos{\chi}\Psi\hat{\mathbf{z}}\times\hat{\mathbf{e}}]. \label{magnetic}
\end{eqnarray} Note that the term $\boldsymbol{\nabla}_{||}V\times\hat{\mathbf{e}}$ corresponds to a magnetic field parallel to the $z$ axis and is subsequently neglected in the paraxial approximation.\\
\indent With these complex $\boldsymbol{\mathcal{E}}$ and $\boldsymbol{\mathcal{B}}$ fields, we can calculate the Poynting vector $\langle \mathbf{S}_\delta(\mathbf{x},t)\rangle=2\textrm{Re}[\boldsymbol{\mathcal{E}}^\ast\times\boldsymbol{\mathcal{B}}]:=\boldsymbol{\mathcal{S}}$, which gives: 
\begin{eqnarray}
\boldsymbol{\mathcal{S}}\simeq\frac{2}{m}\sin^2\chi\textrm{Im}[\Psi^\ast\boldsymbol{\nabla}_{||}\Psi]
\end{eqnarray}
In the same way, we obtain the energy density $u=\varepsilon|\boldsymbol{\mathcal{E}}|^2+|\boldsymbol{\mathcal{B}}|^2\simeq\varepsilon(\sin^2\chi+\cos^2\chi)|\Psi|^2=\varepsilon|\Psi|^2$, which ultimately allows us to deduce the velocity of the energy fluid:
\begin{eqnarray}
\mathbf{v}=\frac{\boldsymbol{\mathcal{S}}}{u}=\frac{2}{m\varepsilon}\sin^2\chi\textrm{Im}[\frac{\boldsymbol{\nabla}_{||}\Psi}{\Psi}]
\end{eqnarray}
Note that this velocity is planar because the field is stationary along z (the neglected term $\boldsymbol{\nabla}_{||}V\times\hat{\mathbf{e}}$  does not contribute to the Poynting vector anyway).   In order to approximate the Bohmian guiding formula, we can average the velocity field along $z$, which gives 
 \begin{eqnarray}
\langle\mathbf{v}\rangle=\frac{\int_0^{D_0}dz u\boldsymbol{\mathbf{v}}}{\int_0^{D_0}dzu}=
\frac{\int_0^{D_0}dz\boldsymbol{\mathcal{S}}}{\int_0^{D_0}dzu}\nonumber\\ \simeq \frac{2}{D_0}\int_0^{D_0}dz\sin^2(\frac{q\pi}{D_0}z)\mathbf{v}_{dBB,||}=\mathbf{v}_{dBB,||}\label{energyflux}
\end{eqnarray}
and allows us to recover the guiding formula as stated in the introduction.\\
\indent This result is the central point of this article. It is important to note once again that our approach here is to consider the Bohmian approach as a practical formalism for understanding light propagation as a fluid. We have started from a classical electromagnetic theory, but similar results would of course be obtainable with a coherent (quasi-classical) quantum field for which it is possible to define a real average electric and magnetic fields  obeying Maxwell's equations \ref{max1}.   The transition from a real field to an imaginary field can be done as we have done here by considering a statistical ensemble or collective of coherent pulses emitted by a laser with a distribution $P(\delta)$ of delay $\delta$. 
Let us note, and we will return to this later, that this way of describing light using Bohmian theory is independent of the more fundamental question of which Bohmian theory is best suited to quantum electrodynamics and photon theory.   This subject has been highly controversial since de Broglie favored a photon particle view \cite{deBroglie1925}, and Bohm favored a photon field approach \cite{Bohm1952b}. The level chosen in this work is that of an effective theory or an alternative modeling of Maxwell's equations, independent of any deeper ontological debates.\\
\indent We believe it is very important to distinguish and deconvolute the two approaches. Particularly in relation to the debate associated with the recent aritcle \cite{Klaers2025}, our approach (alread discussed in \cite{Drezet2025} and further developed here) allows us to simply correct the  errors and conclusions made in \cite{Klaers2025}. Thus, although de Broglie-Bohm theory interpreted as an effective approach for Maxwell's field is not jeopardized by Sharoglazova et al. work, this does not allow us to discuss the true ontological nature of the Bohmian photonic field, which remains a work in progress.  

%%%%%%%%%
\section{Bohmian formalism for a lossy two-dimensional optical microcavity} \label{sec4} 
\subsection{Derivation from Maxwell's equations}
\indent The previous model assumed an  ideal cavity with lossless perfect mirrors. A real experiment, such as the one described in \cite{Klaers2025}, involves a lossy cavity and we must therefore include radiative leakage as well as Ohmic losses  linked to $\varepsilon''$ in our description. Here we still neglect dispersion and suppose a constant permittivity $\varepsilon=\varepsilon'+i\varepsilon''$ with $\varepsilon'>0,\varepsilon''>0$ and $\varepsilon''/\varepsilon'\ll 1$. \\
\indent We consider a one dimensional Fabry-Perot cavity along the $z$ direction and suppose a perfect mirror at $z=0$  and a lossy mirror with reflectivity $r=|r|e^{i\delta}$ at $z=D$. The eigenmode of the cavity with wave-vector $k_z$ obeys the condition
\begin{eqnarray}
1+re^{i2k_zD}=0\label{fresnel}
\end{eqnarray}
which implies 
\begin{eqnarray}
k_z=(q+\frac{1}{2})\frac{\pi}{D} -\frac{\delta}{2D} +i\frac{\ln{|r|}}{2D}
\end{eqnarray} 
In the present case, we are limited to a quasi-ideal cavity and the reflectivity is close to $r=-1$. We set $r=-(1-\eta)$ with $0<\eta\ll 1$ (which implies $\delta=\pm \pi$). We deduce the relation 
\begin{eqnarray}
k_z=\frac{q\pi}{D} +i\frac{\ln{(1-\eta)}}{2D}\simeq  \frac{q\pi}{D} -i\frac{\eta}{2D}
\end{eqnarray}
(here $\delta$ has been absorbed in the definition of $q$) which can be rewritten as 
 \begin{eqnarray}
k_z\simeq  \frac{q\pi}{D} -i\frac{\sqrt{\varepsilon'}\Gamma}{2}
\end{eqnarray} 
 where $\Gamma$ defines a radiative rate with a characteristic time $\tau_R=\Gamma^{-1}$. Note that we have simplified the analysis. In reality, reflectivity $r(\omega,k_{||})$ depends on the angular frequency $\omega$ and the planar wave vector $k_{||}$ (itself linked to $k_z$ by $k_{z}=\sqrt{(\varepsilon\omega^2-k^2_{||})}$ in the different media constituting the cavity and mirrors). Solving Eq.~\ref{fresnel} is an eigenvalue problem  that amounts to finding the poles in the complex plane of the total reflectivity $R$. In general, the solution vectors $k_{||}(\omega)$,  and $k_{z}(\omega)$ are complex numbers that depend on $\omega$. Here we have a highly resonant cavity and we make the approximation $E/m_o\ll 1$, which allows us to write the coefficient $\Gamma(\omega)\simeq \Gamma(m)$, greatly simplifying the analysis.   \\
 \indent In the more general case where the width of the cavity varies slightly with the coordinates $x,y$ in the cavity, we can, as before, introduce an effective mass  $m=\frac{q\pi}{\sqrt{\varepsilon'}D_0}$ and an effective potential $V(x,y)$ such that 
  \begin{eqnarray}
k_z(x,y)\simeq \sqrt{\varepsilon'} (m+V(x,y)-i\frac{\Gamma}{2}).
\end{eqnarray} 
 This amounts to considering a complex potential with a constant dissipative part. The effective Schr\"odinger equation then becomes 
\begin{eqnarray}
i\partial_t\Psi(x,y,t)=-\frac{\boldsymbol{\nabla}_{||}^2}{2\varepsilon' m}\Psi(x,y,t)+(V(x,y)-i\frac{\Gamma}{2}-im\frac{\varepsilon''}{2\varepsilon'})\Psi(x,y,t)
\label{Schrodinger}\end{eqnarray} Note the presence of the Ohmic term leading to the effective 2D potential:
\begin{eqnarray}
V(x,y)\rightarrow V(x,y)-i\frac{\Gamma}{2}-im\frac{\varepsilon''}{2\varepsilon'}
\end{eqnarray}  
When deriving these results, we assume, as in Eq.~\ref{modez}, a function $f_{x,y}(z)$ which is written here as
\begin{eqnarray}
f_{x,y}(z)=\sin{\chi}=\sin{[\sqrt{\varepsilon'}(m+V(x,y)-i\frac{\Gamma}{2})z]}\label{modez2}
\end{eqnarray}
With this expression, the formulas for the electric field $\boldsymbol{\mathcal{E}}(x,yz,t)=\hat{\mathbf{e}}\Psi(x,y,t)e^{-imt}\sin{\chi}$ and the magnetic field Eq.~\ref{magnetic} remain unchanged in the presence of losses (the only difference being the substitutions $\varepsilon\rightarrow \varepsilon'$ and $V\rightarrow V-i\frac{\Gamma}{2}$ without the Ohmic term $-im\frac{\varepsilon''}{2\varepsilon'}$ absent in $k_z$ and $\chi$).\\  
\indent Importantly, due to the imaginary part in Eq.~\ref{modez2}, we recognize a superposition of two growing waves, one in the $+z$ direction $\frac{e^{i\sqrt{\varepsilon'}(m+V(x,y)-i\frac{\Gamma}{2})z}}{2i}$ and one in the $-z$ direction $-\frac{e^{-i\sqrt{\varepsilon'}(m+V(x,y)-i\frac{\Gamma}{2})z}}{2i}$(the real part of the wave vectors is opposite in both cases, which corresponds well to propagation with growing waves: $k'_zk''_z<0$).\\
\indent  Moreover, propagation in the $x-y$ plane as given by Eq.~\ref{Schrodinger} is associated with an absorbing  potential involving the presence of damped evanescent waves (e.g. in the case of a plane wave propagating along a direction $x$ this implies $k'_xk''_x>0$). It is well known that growing waves along $z$ (and decaying waves in the $x-y$ plane) are associated with a phenomenon called leakage radiation which plays a fundamental role in antenna theory and in plasmonics. These waves imply a divergence of the energy transported to infinity in the $\pm z$ direction (here outside the cavity), which is clearly non-physical and stems from the fact that in reality it is a wave emitted by a source located inside the cavity.\\
\begin{figure}
  \centering
  \includegraphics[width=1\textwidth]{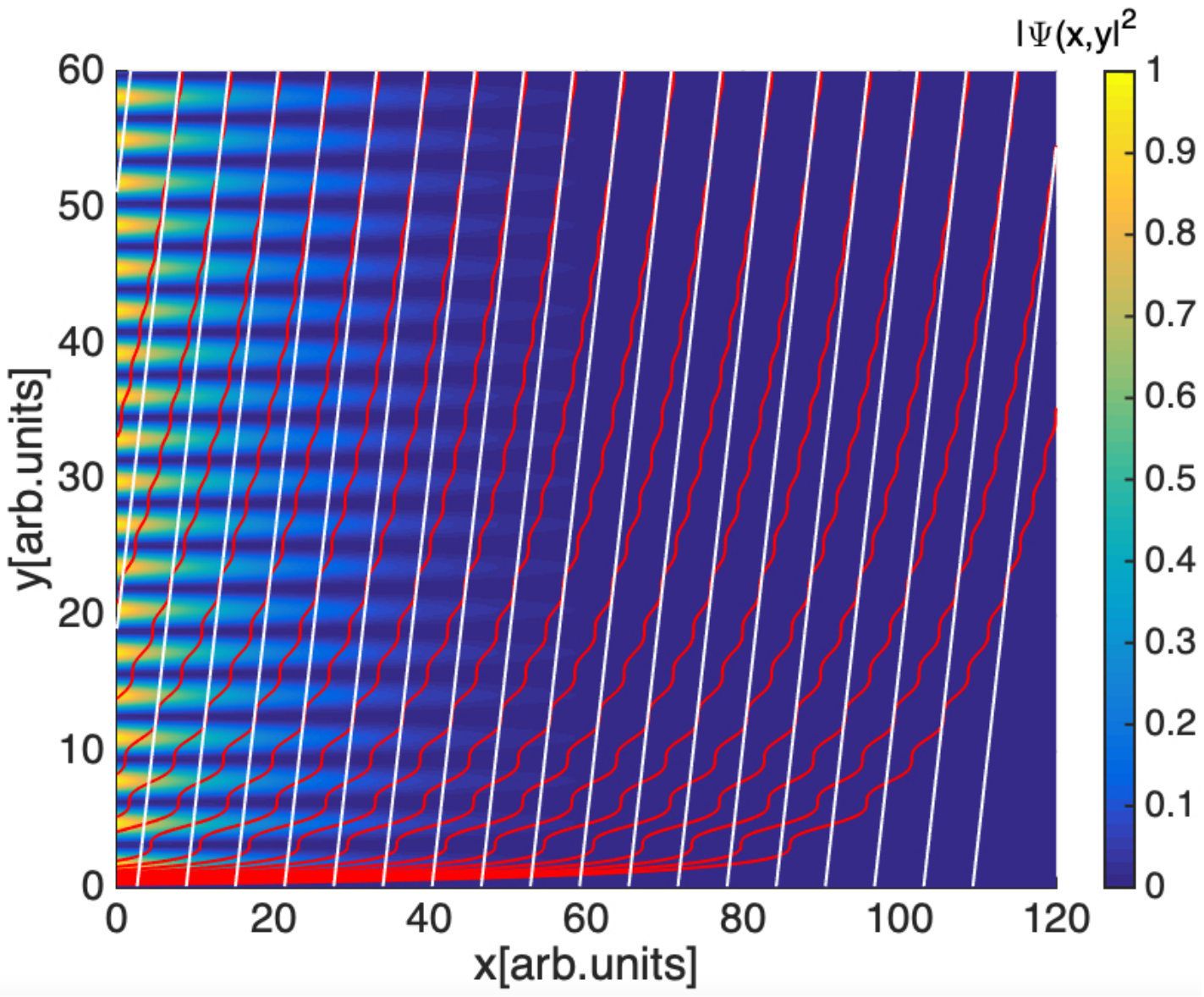}
  \caption{Typical Bohmian trajectories for light inside a lossy optical microcavity. The intensity map  $|\Psi(x,z)|^2$ as a function of $x$ and $z$ shows the decaying of the leaky wave along the $+x$ axis of the planar wave guide and the (very weak) growing of the wave along the $+z$ axis normal to the cavity.  White lines are averaged Bohmian paths and red curves corresponds to actual energy flow curves associated with the local Poynting vector. }
  \label{fig1}
\end{figure}
 \indent Formally, this requires the introduction of a Green's function adapted to the emission of a localized source in the microcavity. The interpretation of a leakage wave must therefore be used with caution and can only be physically interpreted in a limited region of space.\\ 
\indent Moreover It can be rigorously demonstrated that the introduction of finite Green's functions and finite sources into the cavity justifies the use of leaky waves discussed here. In Appendix \ref{appendix}, we show how an antenna located in the microcavity produces a field that asymptotically allows the optical microscope coupled to the cavity to image the leaky waves described in the main body of the article. This justifies the use of this growing leaky wave without having to worry about divergence problems at infinity. More importantly, it justifies the images and observations obtained by Sharoglazova et al. \cite{Klaers2025}.\\   
\indent In what follows, we dont worry anymore about divergences and imaging and we consider only a simple situation, which is of primary interest to us for the purpose of interpreting the experiments reported in \cite{Klaers2025}. We write 
\begin{eqnarray}
\sin{\chi}\simeq \sin{\chi_0}- i\sqrt{\varepsilon'}\frac{\Gamma}{2}z\cos{\chi_0},
\end{eqnarray} 
where we have set $\chi_0=\sqrt{\varepsilon'}(m+V(x,y)z)$, which is justified in the approximation where $\Gamma z\ll 1$ (in the experiment reported in \cite{Klaers2025}, we actually have $\Gamma^{-1}=270 ps$ and $D\simeq 10 \mu m$, which gives $\Gamma D\sim 10^{-5}$, justifying our approximation).\\
\indent This allows us to write the approximate formulas for the Poynting vector: 
\begin{eqnarray}
\boldsymbol{\mathcal{S}}\simeq\frac{2}{m}\sin^2\chi_0\textrm{Im}[\Psi^\ast\boldsymbol{\nabla}_{||}\Psi]-\frac{\sqrt{\varepsilon'}\Gamma}{m}\sin{\chi_0}\cos{\chi_0}|\Psi|^2\hat{\mathbf{z}} +\frac{\varepsilon'\Gamma}{m}(m+V)z\hat{\mathbf{z}}|\Psi|^2\label{a}
\end{eqnarray}
with $\Psi(x,y,t)$ now a solution of Eq.~\ref{Schrodinger} involving $V-i\frac{\Gamma}{2}-im\frac{\varepsilon''}{2\varepsilon'}$ and we still neglected a smaller  term proportional to $\boldsymbol{\nabla}_{||}V\times\hat{\mathbf{e}}$. Similarly, the energy density, and the loss term become:
$u=\varepsilon'|\boldsymbol{\mathcal{E}}|^2+|\boldsymbol{\mathcal{B}}|^2\simeq\varepsilon'|\Psi|^2$ and $\mathcal{W}=2\omega\varepsilon''|\mathcal{E}|^2\simeq 2m\varepsilon'\varepsilon''|\Psi|^2\sin^2\chi_0$.
This fully justifies the use of Bohmian theory for the electromagnetic energy fluid in the cavity, which is the central result of this article. More precisely, the averaged energy velocity is now  given by 
 \begin{eqnarray}
\langle\mathbf{v}\rangle=\frac{\int_0^{D_0}dz u\boldsymbol{\mathbf{v}}}{\int_0^{D_0}dzu}=
\frac{\int_0^{D_0}dz\boldsymbol{\mathcal{S}}}{\int_0^{D_0}dzu}=\frac{1}{\varepsilon' m}\textrm{Im}[\frac{\boldsymbol{\nabla}_{||}\Psi(x,y,t)}{\Psi(x,y,t)}]+ \frac{\Gamma D_0}{2} \hat{\mathbf{z}}\label{b}
\end{eqnarray}
which shows that the planar motion is indeed defined by the Bohmian guiding formula $\mathbf{v}_{dBB,||}$ in accordance with  the assumption of \cite{Drezet2025}. The slow motion along the $+z$ direction with velocity $\frac{D_0}{2\tau_R}$ is characteristic of optical leakage.\\
\indent To illustrate such an effective Bohmian dynamics, Figure \ref{fig1} shows typical trajectories in a microcavity. In the case of a plane wave with $\Psi(x)=e^{ik_x x}$ (see \ref{annwavevector}), we compare the current lines obtained in the $x-z$ plane from formula $\mathbf{v}=\boldsymbol{\mathcal{S}}/u$ (red curves) and the averaged formula \ref{b} (white line). In both cases, the vertical movement corresponds to radiative leaks. Also, planar motion is associated with an energy flow along the $x$ axis.  For the present case (purely theoretical), we have imposed $\varepsilon'=1$, $\varepsilon''=0$, $m=1$ (arbitrary units), $D_0=23\lambda=46\pi/m$, $E/m=10^{-4}$ , $V=0$, and $\Gamma/m=10^{-3}$. 

%%%%%%%%%
\subsection{Generalization for particles with spin-$1/2$}
\indent The analysis in the previous sections was limited to the case of a linear polarization field  $\boldsymbol{\mathcal{E}}=\hat{\mathbf{e}}\Phi(x,y,z)e^{-i\omega t}$ corresponding to the case considered in \cite{Klaers2025,Klaers2023,Drezet2025}. However, it is possible to extend the result a little further to more complex fields written as 
\begin{eqnarray}
\boldsymbol{\mathcal{E}}=[\hat{\mathbf{x}}\Phi_x(x,y,z)+\hat{\mathbf{y}}\Phi_y(x,y,z)]e^{-i\omega t}=[\hat{\mathbf{e}}_+\Phi_+(x,y,z)+\hat{\mathbf{e}}_-\Phi_-(x,y,z)]e^{-i\omega t},
\end{eqnarray}
where $\hat{\mathbf{e}}_\pm=\frac{\hat{\mathbf{x}}\pm i\hat{\mathbf{y}}}{\sqrt{2}}$ denotes left- or right-handed circular polarization unit vectors and where we have $\Phi_\pm(x,y,z)=\Psi_\pm(x,y)f_{x,y}(z)$, $\Phi_{x}(x,y,z)=\Psi_x(x,y)f_{x,y}(z)$, $\Phi_{x}(x,y,z)=\Psi_y(x,y)f_{x,y}(z)$ with $f_{xy}(z)$ given by Eq.~\ref{modez2} and $\Psi_{\pm}(x,y)=\frac{\Psi_x(x,y)\mp i\Psi_y(x,y)}{\sqrt{2}}$ two wave functions forming a Pauli spinor:
\begin{eqnarray}
\mathfrak{F}(x,y)=\left( \begin{array}{c} \Psi_+(x,y)\\ \Psi_-(x,y)\end{array}\right)
\end{eqnarray}
To do this, we use a decomposition obtained by M. Berry \cite{Berry} for the Poynting vector, which is written as 
\begin{eqnarray}
\boldsymbol{\mathcal{S}}\simeq \frac{2}{m}\textrm{Im}[\boldsymbol{\mathcal{E}}^\ast\times(\boldsymbol{\nabla}\times\boldsymbol{\mathcal{E}})]
=\frac{2}{m}\textrm{Im}[\boldsymbol{\mathcal{E}}^\ast(\boldsymbol{\nabla})\boldsymbol{\mathcal{E}}]+\frac{1}{m}\boldsymbol{\nabla}\times\textrm{Im}[\boldsymbol{\mathcal{E}}^\ast\times\boldsymbol{\mathcal{E}}]
\end{eqnarray} with $\boldsymbol{A}(\boldsymbol{\nabla})\boldsymbol{B}=\sum_{k=1}^{k=3}A_k\boldsymbol{\nabla} B_k$.
 This formula contains two contributions, one  $\boldsymbol{\mathcal{S}}_{orb}=\frac{2}{m}\textrm{Im}[\boldsymbol{\mathcal{E}}^\ast(\boldsymbol{\nabla})\boldsymbol{\mathcal{E}}]$ associated with orbital motion and the other $\boldsymbol{\mathcal{S}}_{spin}=\frac{1}{m}\boldsymbol{\nabla}\times\textrm{Im}[\boldsymbol{\mathcal{E}}^\ast\times\boldsymbol{\mathcal{E}}]$ associated with spinorial motion. We deduce the velocity of the energy fluid:
\begin{eqnarray}
\mathbf{v}=\mathbf{v}_{orb}+\mathbf{v}_{spin}
=\frac{\boldsymbol{\mathcal{S}}_{orb}}{u}+\frac{\boldsymbol{\mathcal{S}}_{spin}}{u}
\end{eqnarray}   with $u\simeq\varepsilon'|\boldsymbol{\mathcal{E}}|^2+\frac{1}{m^2}|\boldsymbol{\nabla}\times\boldsymbol{\mathcal{E}}|^2$. Using the Pauli spinor $\mathfrak{F}(x,y)$ we obtain like for Eqs.~\ref{a},\ref{b}
\begin{eqnarray}
\langle\mathbf{v}_{orb}\rangle=
\frac{\int_0^{D_0}dz\boldsymbol{\mathcal{S}}_{orb}}{\int_0^{D_0}dzu}=\frac{1}{m\varepsilon'}\frac{\textrm{Im}[\mathfrak{F}^\dagger\boldsymbol{\nabla}_{||}\mathfrak{F}]}{\mathfrak{F}^\dagger\mathfrak{F}}+ \frac{\Gamma D_0}{2}\hat{\mathbf{z}}\label{c}
\end{eqnarray} 
For the spinorial term, we find at the leading order
\begin{eqnarray}
\boldsymbol{\mathcal{S}}_{spin}\simeq\frac{\sin^2{\chi_0}}{m}\boldsymbol{\nabla}_{||}\times(\mathfrak{F}^\dagger\sigma_z\mathfrak{F}\hat{\mathbf{z}})
\end{eqnarray}
 which implies an average velocity of 
\begin{eqnarray}
\langle\mathbf{v}_{spin}\rangle=
\frac{\int_0^{D_0}dz\boldsymbol{\mathcal{S}}_{spin}}{\int_0^{D_0}dzu}=\frac{1}{2m\varepsilon'}\frac{\boldsymbol{\nabla}_{||}\times(\mathfrak{F}^\dagger\sigma_z\mathfrak{F}\hat{\mathbf{z}})}{\mathfrak{F}^\dagger\mathfrak{F}}
\end{eqnarray}
where $\sigma_z=\left(\begin{array}{cc} 1 &0\\0 &-1\end{array}\right)$ is the Pauli matrix for the $z$ direction.\\
\indent It is remarkable to note that the planar part of this velocity field, i.e., $\langle\mathbf{v}_{orb,||}\rangle+\langle\mathbf{v}_{spin}\rangle=\frac{1}{m\varepsilon'}\frac{\textrm{Im}[\mathfrak{F}^\dagger\boldsymbol{\nabla}_{||}\mathfrak{F}]}{\mathfrak{F}^\dagger\mathfrak{F}}+\frac{1}{2m\varepsilon'}\frac{\boldsymbol{\nabla}_{||}\times(\mathfrak{F}^\dagger\sigma_z\mathfrak{F}\hat{\mathbf{z}})}{\mathfrak{F}^\dagger\mathfrak{F}}$, can be deduced from the Bohmian interpretation of Pauli's equation 
\begin{eqnarray}
i\partial_t\mathfrak{F}(x,y,t)=-\frac{\boldsymbol{\nabla}_{||}^2}{2\varepsilon' m}\mathfrak{F}(x,y,t)+(V(x,y)-i\frac{\Gamma}{2}-im\frac{\varepsilon''}{2\varepsilon'})\mathfrak{F}(x,y,t)
\label{SchrodingerB}\end{eqnarray} 
for the spinor $\mathfrak{F}(x,y,t)$. Indeed, for Eq.~\ref{SchrodingerB} the conservation law for the probability fluid with density $\mathfrak{F}^\dagger\mathfrak{F}$ (see the next subsection for an interpretation of the conservation law) leads to the probability current
 \begin{eqnarray}
 \mathbf{J}_{||}(x,y,t)=\frac{\textrm{Im}[\mathfrak{F}^\dagger\boldsymbol{\nabla}_{||}\mathfrak{F}]}{m\varepsilon'}+\frac{\boldsymbol{\nabla}_{||}\times(\mathfrak{F}^\dagger\sigma_z\mathfrak{F}\hat{\mathbf{z}})}{2m\varepsilon'}
 \end{eqnarray}
which contains a convective and a spinorial contributions. This decomposition is well known and can be deduced from the Dirac equation (Gordon decomposition), which plays an important role in Bohmian mechanics in justifying the correct form of the probability current for a spin-$1/2$ particle \cite{Holland2,Holland3,Dewdney}. It implies the de Broglie-Bohm velocity:
 \begin{eqnarray}
 \mathbf{v}_{dBB,||}(x,y,t)=\frac{\textrm{Im}[\mathfrak{F}^\dagger\boldsymbol{\nabla}_{||}\mathfrak{F}]}{m\varepsilon'\mathfrak{F}^\dagger\mathfrak{F}}+\frac{\boldsymbol{\nabla}_{||}\times(\mathfrak{F}^\dagger\sigma_z\mathfrak{F}\hat{\mathbf{z}})}{2m\varepsilon'\mathfrak{F}^\dagger\mathfrak{F}}\nonumber\\
 =\frac{\textrm{Im}[\Psi_x^\dagger\boldsymbol{\nabla}_{||}\Psi_x+\Psi_y^\dagger\boldsymbol{\nabla}_{||}\Psi_y]}{m\varepsilon'(|\Psi_x|^2+|\Psi_y|^2)}+ \frac{\boldsymbol{\nabla}_{||}\times(\textrm{Im}[\Psi_x^\ast\Psi_y]\hat{\mathbf{z}})}{m\varepsilon'(|\Psi_x|^2+|\Psi_y|^2)}\nonumber\\
 =\frac{\textrm{Im}[\Psi_+^\dagger\boldsymbol{\nabla}_{||}\Psi_++\Psi_-^\dagger\boldsymbol{\nabla}_{||}\Psi_-]}{m\varepsilon'(|\Psi_+|^2+|\Psi_-|^2)}+ \frac{\boldsymbol{\nabla}_{||}\times([|\Psi_+|^2-|\Psi_-|^2]\hat{\mathbf{z}})}{2m\varepsilon'(|\Psi_+|^2+|\Psi_-|^2)}
 \label{vitessePauli}
 \end{eqnarray} which is identical to $\langle\mathbf{v}_{orb,||}\rangle+\langle\mathbf{v}_{spin}\rangle$ deduced from energy conservation. What seems paradoxical, of course, is that the photon (a spin-1 particle by nature) in the paraxial regime behaves here like a massive spin-1/2 particle. However, this is not so surprising if we remember that in this paraxial domain, the polarization fields of light (orthogonal to the direction of propagation $z$) are formally described by SU(2).\\
\indent A few remarks concerning this spinorial formalism are useful here: First, note that in Bohmian theory we can define a local spin vector associated with the motion of the particle \cite{Hiley,HollandB}. This is written as 
\begin{eqnarray}
\boldsymbol{\Sigma}(x,y,t)=\frac{1}{2}\frac{\mathfrak{F}^\dagger\sigma_z\mathfrak{F}}{\mathfrak{F}^\dagger\mathfrak{F}}\hat{\mathbf{z}}=\frac{1}{2}\frac{|\Psi_+|^2-|\Psi_-|^2}{|\Psi_+|^2+|\Psi_-|^2}\hat{\mathbf{z}}=\frac{\textrm{Im}[\Psi_x^\ast\Psi_y]}{|\Psi_x|^2+|\Psi_y|^2}\hat{\mathbf{z}}
\end{eqnarray}
because here the spin has only one component $\Sigma_z(x,y,t)$. In the same context, it should be observed that the role of Berry decomposition \cite{Berry} in Bohmian mechanics for photons (but unrelated to the spinorial formalism used here) had already been considered \cite{Bliokh} in connection with weak-measurement experiments by Kocsis et al. \cite{Kocsis} to measure the Bohmian velocity field of a single photon in a paraxial regime in free space. This important point deserves further scrutiny. We point out, in relation to \cite{Bliokh} that the spin vector $\boldsymbol{\Sigma}$ is connected to the Stokes parameters $s_1, s_2, s_3$, forming a vector $\mathbf{s}$ representing the local polarisation state of light on the Poincar\'e-Bloch sphere ($|\mathbf{s}|=1$). Thus, we have $2\Sigma_z=s_3$.\\
\indent However, the most important observation concerns the transverse nature of the light polarization field. In fact, we did not yet take into account in our analysis (following the phenomenological model of Klaers et al. \cite{Klaers2023,Klaers2025}) the transversality condition $\boldsymbol{\nabla}\cdot\bold{\mathcal{E}}=0$ imposed by Maxwell's equations (condition $\boldsymbol{\nabla}\cdot\boldsymbol{\mathcal{B}}=0$ is directly imposed by the definition $\boldsymbol{\nabla}\times \boldsymbol{\mathcal{E}}\simeq im \boldsymbol{\mathcal{B}}$). Indeed, the constraint $\boldsymbol{\nabla}\cdot\bold{\mathcal{E}}=0$ reads:
\begin{eqnarray}
\partial_x\Psi_x+\partial_y\Psi_y=-\Psi_x\frac{\partial_xf_{x,y}}{f_{x,y}}-\Psi_y\frac{\partial_yf_{x,y}}{f_{x,y}}=-\sqrt{\varepsilon'}\textrm{cotan}{\chi}(\Psi_x\partial_xV+\Psi_y\partial_yV).
\end{eqnarray}
We assume here that we work in the regime where the right hand side is small (i.e., we neglect  $\partial V$) and we write 
 \begin{eqnarray}
\partial_x\Psi_x+\partial_y\Psi_y=\partial_+\Psi_-+\partial_-\Psi_+\simeq 0\label{constraint}
\end{eqnarray}with $\partial_{\pm}=\frac{\partial_x \mp i \partial_y}{\sqrt{2}}=\frac{\partial}{\partial x_\mp} $ and $x_\pm=\frac{x \mp i y}{\sqrt{2}}$. The Pauli Eq.~\ref{SchrodingerB} and constraint \ref{constraint} must therefore be solved consistently, which for a general problem does not admit any trivial solutions a priori. However, we can approximately solve the problem by introducing a stream function $Q(x,y,t)$ such that we have 
 \begin{eqnarray}
\Psi_x=-\partial_yQ,& \Psi_y=\partial_xQ\nonumber\\
\Psi_+=-i\partial_+Q,& \Psi_-=i\partial_-Q
\end{eqnarray}
 which automatically satisfies Eq.~\ref{constraint}. In order to satisfy Pauli's equations \ref{SchrodingerB}, we impose Schrödinger's equation for $Q$: 
 \begin{eqnarray}
i\partial_tQ(x,y,t)=-\frac{\boldsymbol{\nabla}_{||}^2}{2\varepsilon' m}Q(x,y,t)+(V(x,y)-i\frac{\Gamma}{2}-im\frac{\varepsilon''}{2\varepsilon'})Q(x,y,t)
\label{SchrodingerC}\end{eqnarray} 
 This condition allows us to satisfy Eq.~\ref{SchrodingerB} with the same degree of precision as for Eq.~\ref{constraint} (i.e., by neglecting  $\partial V$).\\
\indent With this description, the Bohmian velocity Eq.~\ref{vitessePauli} is written as: 
\begin{eqnarray}
 \mathbf{v}_{dBB,||}(x,y,t) =\frac{\textrm{Im}[\partial_xQ^\dagger\boldsymbol{\nabla}_{||}\partial_xQ+\partial_yQ^\dagger\boldsymbol{\nabla}_{||}\partial_yQ]}{m\varepsilon'(|\partial_xQ|^2+|\partial_yQ|^2)}+ \frac{\boldsymbol{\nabla}_{||}\times(\textrm{Im}[\partial_xQ^\ast\partial_yQ]\hat{\mathbf{z}})}{m\varepsilon'(|\partial_xQ|^2+|\partial_yQ|^2)}\nonumber\\
 =\frac{\textrm{Im}[\partial_+Q^\dagger\boldsymbol{\nabla}_{||}\partial_+Q+\partial_-Q^\dagger\boldsymbol{\nabla}_{||}\partial_-Q]}{m\varepsilon'(|\partial_+Q|^2+|\partial_-Q|^2)}+ \frac{\boldsymbol{\nabla}_{||}\times([|\partial_+Q|^2-|\partial_-Q|^2]\hat{\mathbf{z}})}{2m\varepsilon'(|\partial_+Q|^2+|\partial_-Q|^2)}
 \label{vitessePauliB}
 \end{eqnarray}
 Strictly speaking, this dynamic must play a role in the analysis of the experiment detailed in \cite{Klaers2025} in order to go beyond the phenomenological approach based on a single wave $\Psi$ \cite{Klaers2023} and neglecting the role of polarization. A more detailed and precise analysis of the experiments reported \cite{Klaers2025}  will be published at a later date. However, it should be noted that this does not alter the general conclusions of our critique presented in \cite{Drezet2025}. In particular, without going into detail, the Bohmian velocity $v_x$ along the waveguides studied in \cite{Klaers2025} (i.e., in the evanescent regime) is still quantitatively dominated by  $v_x\simeq \frac{\Gamma}{2\sqrt{ (-2m\Delta)}}\simeq 30 km/s$  (where $\Delta\simeq -0.04 meV$ is an energy detuning \cite{Klaers2025}) in agreement with our previous analysis \cite{Drezet2025}. 
%%%%%%%%%%%%%%%%
%%%%%%%%%%%%%%%%
\subsection{Conservation of probability and stochastic Bohmian mechanics}
\indent It is crucial to note that the effective Schrodinger equation \ref{Schrodinger} is non-unitary due to the presence of the complex potential $V-i\frac{\Gamma}{2}-im\frac{\varepsilon''}{2\varepsilon'}$.   In particular, the local conservation law for the electron fluid becomes
\begin{eqnarray}
\partial_t|\Psi|^2+\boldsymbol{\nabla}_{||}(|\Psi|^2\mathbf{v}_{dBB,||})+(\Gamma+m\frac{\varepsilon''}{\varepsilon'})|\Psi|^2=0\label{energyconser}
\end{eqnarray} This equation can be derived directly from Eq.~\ref{Schrodinger} or from the local energy conservation condition \ref{conse} by averaging over $z$: $\partial_t \int_0^{D_0}dz u=-\int_0^{D_0}dz\boldsymbol{\nabla}\cdot \boldsymbol{\mathcal{S}}-\int_0^{D_0}dz \mathcal{W}$. A similar equation is deduced from Eq.~\ref{SchrodingerB} with the density $\mathfrak{F}^\dagger\mathfrak{F}$ replacing $|\Psi|^2$ and with the spin-dependent velocity fields Eq.~\ref{vitessePauli}.  We stress that Eq.~\ref{energyconser} involves a radiative loss term $\Gamma|\Psi|^2$ that must be interpreted as an effective absorption. It is indeed effective in the sense that it exists only from a perspective where we neglect the presence of the third dimension $z$ characterized by a complex wave vector $k_z=\sqrt{\varepsilon'}(m+V(x,y)-i\frac{\Gamma}{2})$ appearing in Eq.~\ref{modez2}. The second loss term $m\frac{\varepsilon''}{\varepsilon'}|\Psi|^2$ is associated with internal absorption due to Ohmic dissipation. Clearly, if we are only considering the 2D motion the two channels associated with radiative and Ohmic losses are indistinguishable. This is reminiscent of  works done in quantum optics modeling dissipation $\varepsilon''$ as effective optical channels associated with leakage \cite{Loudon,Tame}.\\
\indent From a Bohmian perspective, there is a strong analogy between the present model and the work of Bell and D\"urr et al. in QFT based on a stochastic version of de Broglie Bohm theory \cite{Bell,Sudbery,Vink,Tumulka} (see also \cite{Drezet} for a link to measurement theory and the notion of arrival time \cite{Dass,GTZ} in Bohmian mechanics). To be more explicit would require a detailed analysis, which we will briefly outline in this following.\\
\indent First, the conservation equation \ref{energyconser} characterizes a master equation for an electromagnetic fluid. The quantity $u_{2D}=\varepsilon' D_0|\Psi|^2$ defines a surface energy density for this fluid, and $(\Gamma+m\frac{\varepsilon''}{\varepsilon'})u_{2D}$ quantifies the rate of disappearance of this fluid through Ohmic absorption and radiative leakage channels. 
In order to obtain a better Bohmian analogy, it must be possible to talk about the probability of the photon's presence in the 2D cavity, but this controversial notion is not universally accepted in the quantum optics community (see e.g. \cite{cook,Sipe,Birula} for interesting discussions and proposals). However, in the present case, the photon is an effective particle with mass $\omega\simeq m$, and it is a priori justified to introduce a particle surface density $\rho_{2D}(x,y,t)=u_{2D}(x,y,t)/\omega\simeq u_{2D}(x,y,t)/m$.\\ \indent Furthermore, although the analysis in this work is essentially classical, it can easily be extended to the quantum domain. Thus, if we are limited to a single photon, the electromagnetic field of the effective particle in the cavity can be deduced from the second-quantized formalism of QED, and we can set 
\begin{eqnarray}
\boldsymbol{\mathcal{E}}(\mathbf{x},t):=\langle 0|\hat{\mathbf{E}}^{(+)}(\mathbf{x},t)|1 \rangle\nonumber\\
\boldsymbol{\mathcal{B}}(\mathbf{x},t):=\langle 0|\hat{\mathbf{B}}^{(+)}(\mathbf{x},t)|1 \rangle\label{QED}
\end{eqnarray}
where $|0\rangle$, $|1\rangle$ denote a Fock state  with zero or 1 photon respectively,  and where $\hat{\mathbf{E}}^{(+)}$  and $\hat{\mathbf{B}}^{(+)}$ are the positive frequency parts of the quantum electromagnetic field operators.    The fields  \ref{QED} appear  in the transition amplitudes during photon emission and absorption phenomena and are associated in \cite{Sipe,Drezet2005} with real electromagnetic wave functions obeying classical Maxwell's equations \ref{max2} in the optical environment under consideration (see also \cite{Huttner,Gruner,Drezet2017} for a rigorous discussion in a lossy and dispersive dielectric environment).\\
\indent If we limit ourselves to a phenomenological description, the photon density $\rho_{2D}(x,y,t)$ is associated with the second quantized wave function $\Psi(x,y,t):=\langle 0|\hat{\Psi}^{(+)}(x,y,t)|1 \rangle$ with
\begin{eqnarray}
\rho_{2D}(x,y,t)\simeq\frac{\varepsilon' D_0}{m}|\Psi(x,y,t)|^2
\end{eqnarray} In order to normalize this wave function in a lossy cavity, we can start from a wave packet defined in the distant past with an initial energy $\omega\simeq m$, which allows us to introduce a normalized wave function $\tilde{\Psi}(x,y,t)=\sqrt{(\frac{m}{\varepsilon' D_0})}\Psi(x,y,t)$ at initial time $t_0$ (i.e., $\int dxdy |\tilde{\Psi}(x,y,t_0)|^2=1$). The effective Schr\"odinger equation \ref{Schrodinger} for this wave function $\tilde{\Psi}(x,y,t)$ therefore describes a non-unitary evolution associated with a non-Hermitian Hamiltonian $\hat{H}=-\frac{\boldsymbol{\nabla}_{||}^2}{2\varepsilon' m}+V(x,y)-i\frac{\Gamma}{2}-im\frac{\varepsilon''}{2\varepsilon'}$  due to absorption.\\ 
\indent According to the formalism introduced by Bell \cite{Bell,Sudbery,Vink,Tumulka}, the law of conservation of probability fluid can be written in general terms as 
\begin{eqnarray}
\partial_t|\tilde{\Psi}(x,y,t)|^2=2\textrm{Im}[\tilde{\Psi}^\ast(x,y,t)\hat{H}\tilde{\Psi}(x,y,t)]
\end{eqnarray}
 in which we can distinguish between a Hermitian contribution $\hat{H}^{(0)}=-\frac{\boldsymbol{\nabla}_{||}^2}{2\varepsilon' m}+V(x,y)$ giving rise to $2\textrm{Im}[\tilde{\Psi}^\ast\hat{H}^{(0)}\tilde{\Psi}]=-\boldsymbol{\nabla}_{||}(|\tilde{\Psi}|^2\mathbf{v}_{dBB,||})$  associated with the deterministic evolution, and a non-Hermitian contribution $\hat{H}^{(L)}=-i\frac{\Gamma}{2}-im\frac{\varepsilon''}{2\varepsilon'}$ leading  to $2\textrm{Im}[\tilde{\Psi}^\ast\hat{H}^{(L)}\tilde{\Psi}]=-(\Gamma+m\frac{\varepsilon''}{\varepsilon'})|\tilde{\Psi}|^2$ and associated  with losses (similar expressions are easily obtained for the spin-dependent Pauli equation).  According to Bell, if $\hat{H}$ were purely Hermitian we could introduce a probability of quantum jumps (or Bell jumps) using the formula 
\begin{eqnarray}
\sigma_t(x,y|x',y'):=\frac{\mathcal{J}_{x,y;x',y',t}^+}{|\tilde{\Psi}(x',y',t)|^2}\nonumber\\
\end{eqnarray} with $\mathcal{J}_{x,y;x',y',t}=2\textrm{Im}[\tilde{\Psi}^\ast(x,y,t)\mathcal{H}_{x,y;x',y'}\tilde{\Psi}(x',y',t)]$, $\mathcal{H}_{x,y;x',y'}:=\langle x,y|\hat{H}|x',y'\rangle$ and 
where the plus sign indicates that this probability is taken to be zero if $\mathcal{J}_{x,y;x',y',t}$ is negative. For such a Hermitian Hamiltonian we would have  $\mathcal{H}_{x,y;x',y'}=\mathcal{H}_{x',y';x,y}^\ast$, $\mathcal{J}_{x,y;x',y',t}=-\mathcal{J}_{x',y';x,y,t}$ and this would imply a master equation with both source terms and loss terms: 
\begin{eqnarray}
\partial_t|\tilde{\Psi}(x,y,t)|^2=\int dx'dy'[\sigma_t(x,y|x',y')|\tilde{\Psi}(x',y',t)|^2-\sigma_t(x',y'|x,y)|\tilde{\Psi}(x,y,t)|^2].\label{master}
\end{eqnarray} From such a master equation we have obviously probability conservation $\frac{d}{dt}\int dxdy|\tilde{\Psi}(x,y,t)|^2=0$. 
However, here with the non Hermitian operator $\hat{H}^{(L)}$ we must change the definition  of Bell jumps and assume the most general form including source terms $S$ and loss terms $L$:
\begin{eqnarray}
\sigma^{(S)}_t(x,y|x',y'):=\frac{\mathcal{J}_{x,y;x',y',t}}{|\tilde{\Psi}(x',y',t)|^2},&\sigma^{(S)}_t(x',y'|x,y)=0 \nonumber\\ \textrm{if } \mathcal{J}_{x,y;x',y',t}\geq 0\nonumber\\
\sigma^{(L)}_t(x',y'|x,y):=-\frac{\mathcal{J}_{x,y;x',y',t}}{|\tilde{\Psi}(x,y,t)|^2}, &\sigma^{(L)}_t(x,y|x',y')=0  \nonumber\\ \textrm{if } \mathcal{J}_{x,y;x',y',t}\leq 0
\end{eqnarray} which again leads to a master equation:
\begin{eqnarray}
\partial_t|\tilde{\Psi}(x,y,t)|^2=\int dx'dy'[\sigma^{(S)}_t(x,y|x',y')|\tilde{\Psi}(x',y',t)|^2-\sigma^{(L)}_t(x',y'|x,y)|\tilde{\Psi}(x,y,t)|^2].\nonumber\\ \label{master}
\end{eqnarray}
Here, with the separation $\hat{H}^{(0)}$ and $\hat{H}^{(L)}$, we have only a loss term $L$  and no source term $S$ and we get 
\begin{eqnarray}
\partial_t|\tilde{\Psi}(x,y,t)|^2+\boldsymbol{\nabla}_{||}(|\tilde{\Psi}|^2\mathbf{v}_{dBB,||})=-\int dx'dy'\sigma^{(L)}(x',y'|x,y)|\tilde{\Psi}(x,y,t)|^2.\label{master}
\end{eqnarray}
where $\sigma^{(L)}(x',y'|x,y)=(\Gamma+m\frac{\varepsilon''}{\varepsilon'})\delta(x-x')\delta(y-y')$. The presence of the Dirac distribution implies  that this quantum jump process is locally defined and occurs only if $x=x'$ and $y=y'$. Physically, Bell's jump corresponds here to the disappearance of a particle at point $x,y$ associated with absorption. Note that we have $\frac{d}{dt}\int dxdy|\tilde{\Psi}(x,y,t)|^2=-(\Gamma+m\frac{\varepsilon''}{\varepsilon'})\int dxdy|\tilde{\Psi}(x,y,t)|^2$ implying an exponential decay $\int dxdy|\tilde{\Psi}(x,y,t)|^2=e^{-(\Gamma+m\frac{\varepsilon''}{\varepsilon'})(t-t_0)}$ of the photon probability in the cavity. Such a Bell-type stochastic dynamic (based on a non-Hermitian Hamiltonian) implies that the effective 2D photon trajectories in the planar microcavity can be interrupted at a rate $(\Gamma+m\frac{\varepsilon''}{\varepsilon'})|\tilde{\Psi}|^2$ per unit time and surface area.         \\
\indent A few remarks on the significance of our stochastic model are useful here:\\
1) First, it is not really necessary to consider the stochastic approach in order to interpret the experiments described in \cite{Klaers2025} and criticized in \cite{Drezet2025}.  In fact, the Ohmic losses proportional to $\varepsilon''$ are negligible here, and only the radiative losses related to $\Gamma$ remain. The stochastic interpretation is only useful in effective 2D modeling that forgets the physical nature of radiative losses (leakage) as a photonic movement along the third dimension $z$ and directed toward the outside of the optical cavity.\\ 
2) This contrasts sharply with other analyses \cite{Nikolic,Wang,Dickau,Klaersarxiv}, which have attempted to respond to Sharoglazova et al.\cite{Klaers2025}, and which rely on an intrinsically stochastic modified Bohmian dynamics  in order  to  describe the passage and propagation of photons between the waveguides considered in \cite{Klaers2025}. As we have shown in \cite{Drezet2025}, a completely deterministic Bohmian dynamics is sufficient to account for the results of \cite{Klaers2025} in their entirety, contrary to what is claimed by the authors of \cite{Nikolic,Wang,Dickau,Klaersarxiv}. The results obtained here confirm our previous analysis \cite{Drezet2025} and, as indicated above, radiative losses at a rate $\Gamma$ are a three-dimensional deterministic process that can nevertheless effectively be analyzed stochastically when viewed from the $x-y$ plane where Bohmian dynamics are confined.

%%%%%%%%%%%              
\section{Final remarks and conclusion}\label{sec5}
\indent To conclude this work it is important to reaxamine the context in which the present analysis is done.
We remind that the article by Sharoglazova et al. \cite{Klaers2025}  has sparked significant attention in both the scientific community \cite{Drezet2025,Nikolic,Wang,Dickau,Klaersarxiv} (see also \cite{Sienicki,Daem,Waegell}) and the media \cite{news1,news2,news3,news4}, due to its striking claim that a micro-photonics experiment challenges—or even refutes—the de Broglie–Bohm interpretation of quantum mechanics.\\
\indent Their experiment uses evanescent waves confined in parallel optical waveguides, in a regime where Maxwell’s equations are formally equivalent to an effective Schrödinger equation—thus enabling a photonic analog of 2D Bohmian mechanics. Their central claim is that while Bohmian mechanics predicts zero particle velocity in evanescent waves, their observations show light ``hopping'' between guides, and they even propose an experimental reconstruction of a velocity field that differs from the Bohmian one (their velocity   exists in evanescent fields). They argue that this reveals a fundamental contradiction with Bohmian theory, which should therefore be revised or extended.\\
\indent In our response \cite{Drezet2025}, we have shown that this interpretation is incorrect and that the issue stems mostly from a mischaracterization of radiative (i.e., leaky) waves (a second contribution is associated with the finite temporal witdh of the excitation but is quantitatively much smaller). The authors of \cite{Klaers2025} perceive a contradiction in `observing' photon motion where Bohmian theory predicts none. They introduce a different (i.e., non Bohmian) definition of speed which is not vanishing in a pure evanescent  field.  However, the real question they should have asked themselves, but didn’t,  is: why assume that a velocity must be assigned to photons in such a waveguide configuration, especially in the evanescent regime?\\
\indent As we have shown in \cite{Drezet2025} the existence of a velocity in a evanescent field arises because the camera records a signal: light couples from the pump laser to a molecular medium, emits photons into the waveguides, and these are finally imaged. This suggests motion—hence the need for a velocity field, even in an evanescent context.
Moreover, since Bohmian velocity cancels out in an ideal evanescent field, this suggests a contradiction with de Broglie-Bohm theory, which is why Sharoglazova et al. \cite{Klaers2025} believe they have refuted Bohmian theory. As a solution or alternative the authors of \cite{Klaers2025} introduce another operational definition of velocity that can be deduced from the populations of photons observed in their optical waveguides.\\ 
\indent However, the most natural and general definition should be based on the energy flux passing through the waveguides in the microcavity. Indeed, the observation of photons on the camera is directly related to electromagnetic energy transported through the optical system.  Moreover, and therein lies their fundamental oversight,  Bohmian mechanics here is only an effective framework derived from Maxwell’s equations. A non-zero Bohmian velocity is directly linked to a non-zero Poynting vector. The present work clearly justifies this point since Bohmian mechanics is actually an alternative hydrodynamic formulation of Maxwell's theory in the microcavity. The very fact that we detect light associated with evanescent waves implies, through continuity, that the Bohmian velocity cannot be strictly zero inside the cavity. The wave cannot be purely evanescent: This looks like a paradox.\\
 \indent In \cite{Drezet2025} we demonstrated through detailed analysis that the observed signals necessarily involve optical losses—in particular, leakage from the microcavity—which are essential for the imaging process itself. These losses give rise to a finite Poynting vector and hence a non-zero Bohmian velocity inside the cavity. The new analysis provided here confirms this finding and gives a firm foundation to the Bohmian analogy. This resolves the apparent paradox  and fully agrees with all data \cite{Klaers2025} without altering or weakening Bohmian theory.\\
\indent A technical point that we developed in Appendix \ref{appendix} concerns the link between leaky waves, which are by definition divergent, and the finite far-field imaged by a microscope detecting radiative leaks from the cavity. We have shown (drawing on classical work \cite{sommerfeld,russe,DrezetLRM,Nikitin,Berthel}) that leaky waves leave their fingerprints in the optical far-field and do indeed provide a physical understanding of the propagation processes involved in optical microcavities.\\
 \indent Far beyond this debate surrounding the interpretation of \cite{Klaers2025}, the formalism developed here offers a new and interesting perspective on Bohmian mechanics applied to light, which is worth summarizing. First, we started with an effective description of the classical Maxwell equations. The Bohmian formalism is constructed as an approximation, just like the effective Schrodinger equation deduced in the paraxial regime. The most important point is that we can connect the velocity deduced from the energy flow to the Bohmian velocity of the Schrodinger fluid.  This formal construction is therefore independent of the more fundamental debates concerning Bohmian ontology applied to QED and the photon. Of course, the formalism can also be applied to a single photon, and the link between the two issues reappears, but this will not be analyzed in this work. Three important consequences of our work deserve further analysis: i) The problem of spin associated with our description of an optical field (traditionally spin 1) by an effective spin-$1/2$ field connected to the Pauli equation). ii) In addition to the simulation of quantum analog 2D massive spin-$1/2$ Bohmian particles in complex potential,  it could be interesting to see if this method of averaging several pairs of real fields with random phases into a single complex field has deeper meaning and if it helps explain why complex fields appear so naturally in quantum theory (this issue is related to the recent controversy \cite{Gisin,rebut,rebut2}). iii) The possibility of supplementing the deterministic description with a Bell-style stochastic approach, which allows a link to be made with modern QFT and attempts to describe a Bohmian theory with a variable number of particles \cite{Tumulka}. We believe that all of these results can be taken as a source of inspiration for future research. \\
\indent From a broader perspective, we believe that the hydrodynamic analogy justified here would allow us to analyze many two-dimensional optical phenomena in optical cavities using the concept of Bohmian trajectories. Of course, current works are still limited and approximate because they neglect the polarization and spin effects associated with the E and B fields. However, as we have shown  a more general effective Bohmian dynamics based on Pauli's equation for a spin-$1/2$ particle must actually applies in a planar optical micro cavity.  By analogy with the Dirac equation \cite{Hiley,HollandB}, it should be possible to reveal contributions related to polarization in the energy current and in the Bohmian velocity (this is strongly related to recent questions concerning arrival-time for electrons in Bohmian mechanics \cite{Dass,Drezet,GTZ}). This opens up interesting prospects for this work in connection with developments in nanophotonics involving highly polarized surface states such as topological and chiral systems.\\
%%%%%%%
\acknowledgments{ We would like to thank Dustin Lazarovici for his analysis and comments, as well as for his support in this work. AD would like to dedicate this work to the memory of Basil Hiley, who was David Bohm's close collaborator and a great fighter against  quantum orthodoxy.} 
%%%%%%%%
\appendix
\section{Leakage radiation and finite source}\label{appendix}
\indent Consider a current distribution $\boldsymbol{\mathcal{E}}(\mathbf{x},t)$ localized inside the lossy cavity in the vicinity of the mirror $z=0$ and such that the electric field obeys the wave equation (derivable from Maxwell's equations in a polarizable dielectric medium): 
\begin{eqnarray}
-\varepsilon\partial_t^2\boldsymbol{\mathcal{E}}+\boldsymbol{\nabla}^2\boldsymbol{\mathcal{E}}=\partial_t\boldsymbol{\mathcal{J}}.
\end{eqnarray}
Here we will take as a simple example a singular line current distribution such that $\boldsymbol{\mathcal{J}}= I_0\hat{\mathbf{y}}\delta(z-d)\delta(x)e^{-i\omega t}$ and $I_0$ an electric current. We are looking for a solution of the form $\boldsymbol{\mathcal{E}}=\hat{\mathbf{y}}\Phi(x,z)e^{-i\omega t}$ and therefore we have  
\begin{eqnarray}
\varepsilon\omega^2\Phi+(\partial_x^2+\partial_z^2)\Phi=-i\omega I_0\delta(z-d)\delta(x).
\end{eqnarray}
The solution is written as a Fourier integral $\Phi(x,z)=\int_{-\infty}^{+\infty} dk_x\tilde{\Phi}(k_x,z)e^{ik_x x}$ and if $d\ll D_0$  we deduce:
\begin{eqnarray}
\tilde{\Phi}(k_x,z)\simeq\frac{i\omega I_0}{2\pi k_z}\sin{(k_z z)}& \textrm{if } z\in[0,d]\nonumber\\
\tilde{\Phi}(k_x,z)\simeq\frac{i\omega I_0d}{2\pi }\frac{e^{ik_zz}+re^{2ik_zD}e^{-ik_zz}}{1+re^{2ik_zD}}& \textrm{if } z\in[d,D]\nonumber\\
\tilde{\Phi}(k_x,z)\simeq\frac{i\omega I_0d}{2\pi }\frac{t e^{ik_zz}}{1+re^{2ik_zD}}& \textrm{if } z\geq D
\end{eqnarray}
To simplify we suppose the same permitivity outside the cavity.
These cumbersome (Sommerfeld) integrals are asymptotically evaluated using  a closed contour in the complex plane and by taking the complex variable $k_x=\omega\sqrt{\varepsilon}\sin{\xi}$ with $\xi=\xi'+i\xi'' \in \mathbb{C}$ (for a discussion of the method see \cite{sommerfeld,russe,DrezetLRM}). We emphasize that choosing the same permittivity inside and outside the cavity simplifies the analysis by eliminating additional branch cuts in the complex plane that would give rise to the presence of Norton lateral waves propagating along the walls of the microcavity and decaying as $\sim 1/|x|^{3/2}$~\cite{Nikitin,russe,DrezetLRM}.\\  
\indent i) First, for $z\geq d$ there is a singular solution $\Phi_P(x,z)$ associated with the reflectivity pole  $\xi_P$ solution of the condition $1+r_Pe^{2ik_{zP}D}=0$ (see Eq.~\ref{fresnel}) with $k_{xP}=\omega\sqrt{\varepsilon}\sin{\xi_P}$ and $k_{zP}=\omega\sqrt{\varepsilon}\cos{\xi_P}$ the eigenvalues solution of Eq.~\ref{fresnel} ($r_P,t_P$ are the Fresnel coeeficients corresponding to such eigenvalues). Note that Eq.~\ref{fresnel} admits the solutions $\pm\xi_P$  but only $\xi_P$ contributes to the complex integration. This singular solution $\Phi_P(x,z)$ reads 
\begin{eqnarray}
\Phi_P(x,z)=\frac{-\omega I_0de^{ik_{xP}|x|}}{\frac{\partial(re^{2ik_zD})}{\partial k_x}|_{k_{xP}}}f_P(z)\label{pole}
\end{eqnarray}
with 
\begin{eqnarray}
f_P(z)\simeq e^{ik_{zP}z}\Theta(|\varphi-\xi'_P|)\nonumber\\+r_Pe^{2ik_{zP}D}e^{-ik_{zP}z}\Theta(|\bar{\varphi}-\xi'_P|)& \textrm{if } z\in[d,D]\nonumber\\
 f_P(z)=t_Pe^{ik_{zP}z}\Theta(|\varphi-\xi'_P|)& \textrm{if } z\geq[D]
\end{eqnarray} and where the presence of two Heaviside functions must be noted. These functions have as arguments the angles $\varphi$ and $\bar{\varphi}$ corresponding to the two cylindrical coordinate systems  $[x=\rho\sin{\varphi},z=\rho\cos{\varphi}]$ (with $\varphi\in[-\pi/2,+\pi/2]$) and $[x=\bar{\rho}\sin{\bar{\varphi}}, D-z=\bar{\rho}\cos{\bar{\varphi}}]$ (with $\bar{\varphi}\in[-\pi/2,+\pi/2]$)adapted to the two mirrors located respectively at  $z=0$ and $z=D$.
$\Phi_P(x,z)$ corresponds to the leaky wave analyzed in the body of the article. Due to the presence of the Heaviside functions, $\Phi_P(x,z)$ only exists beyond critical angles associated with $\pm\xi'_p$ i.e., for $|\varphi|\geq \xi'_P$ and $|\bar{\varphi}|\geq \xi'_P$.\\
\indent This solution therefore has angular discontinuities, which makes it useful only in the asymptotic region far from the source . Furthermore, in the paraxial domain considered here, we have $\omega=m+E$ with $E/m\ll 1$, so the solution $\xi_P$ can be approximated as: 
\begin{eqnarray}
\xi_P\simeq\sqrt{\frac{2E}{m}}+i\frac{\Gamma+m\frac{\varepsilon''}{\varepsilon'}}{2\sqrt{(2Em)}}
\end{eqnarray} which gives the wave vectors 
\begin{eqnarray}
k_{xP}\simeq \sqrt{(2m\varepsilon' E)}+i\frac{m\varepsilon'(\Gamma+m\frac{\varepsilon''}{\varepsilon'})}{2\sqrt{(2m\varepsilon' E)}}\nonumber\\
k_{zP}\simeq \sqrt{\varepsilon'}(m-i\frac{\Gamma}{2})\label{annwavevector}
\end{eqnarray} 
in agreement with the discussion in the main text. Since  the angle $\xi'_P\simeq \sqrt{(2E/m)}\ll 1$,  the solution $\Phi_P(x,z)$ can in fact be used in a wide range of the optical microcavity avoiding the surrounding of the $x=0$ plane. It is also important to note that since this polar solution is not valid in the angular domain $|\varphi|\leq \xi'_P$, it does not actually diverge to infinity.  Thus, in the vicinity of $|\varphi|\gtrsim\xi'_P=\sqrt{2E/m}$, the norm $|\Phi_{P}| \sim  e^{-\sqrt{\varepsilon'}\rho(\Gamma+m\frac{\varepsilon'‘}{\varepsilon’})(\theta/\xi'_P-1)}$ tends toward zero if $\rho $ increases toward infinity.
\\  
\indent ii) The second contribution $\Phi_{SDP}(x,z)$ comes from the steepest descent path (SDP) and allows us to evaluate the far-field emitted by the current distribution \cite{russe,DrezetLRM}. Indeed, in the domain $z\in [0,d]$ the SDP solution reads: 
$\Phi_{SDP}=\frac{\omega I_0}{4\pi}[H_0^{(+)}(\omega\sqrt{\varepsilon}\sqrt{(x^2+z^2)})-e^{-i\omega\sqrt{\varepsilon}\cos{\beta}d}H_0^{(+)}(\omega\sqrt{\varepsilon}\sqrt{(x^2+(d-z)z^2)})]$ with $\cos{\beta}=\frac{d-z}{\sqrt{(x^2+(d-z)z^2)}}$. This contribution tends to vanish in the limit $d\rightarrow 0^+$. Moreover, in the domain $z\geq d$, the SDP solution is approximately written as 
\begin{eqnarray}
\Phi_{SDP}(x,z)\simeq\omega\sqrt{\varepsilon}\tilde{\Psi}_{c}(\omega\sqrt{\varepsilon}\sin{\varphi})\cos{\varphi}H_0^{(+)}(\omega\sqrt{\varepsilon}\rho)\nonumber\\ +\omega\sqrt{\varepsilon}\tilde{\Psi}_{c}(\omega\sqrt{\varepsilon}\sin{\bar{\varphi}})\cos{\bar{\varphi}}H_0^{(+)}(\omega\sqrt{\varepsilon}\bar{\rho})r_{\bar{\varphi}} e^{i\omega\sqrt{\varepsilon}\cos{\bar{\varphi}}D}\nonumber\\ \textrm{if } z\in[d,D]\nonumber\\
\Phi_{SDP}(x,z)\simeq\omega\sqrt{\varepsilon}\tilde{\Psi}_{c}(\omega\sqrt{\varepsilon}\sin{\varphi})\cos{\varphi}H_0^{(+)}(\omega\sqrt{\varepsilon}\rho)t_\varphi \nonumber\\ \textrm{if } z\geq[D]\nonumber\\
\end{eqnarray}
where we have introduced the  Fourier transform for the cavity field $\tilde{\Psi}_{c}(k_x)=\frac{i\omega I_0d}{2\pi }\frac{1}{1+re^{2ik_zD}}$. The Hankel function is written asymptotically here as $H_0^{(+)}(\omega\sqrt{\varepsilon}\rho)\simeq e^{-i\pi/4}e^{i\omega\sqrt{\varepsilon}\rho}\sqrt{\frac{2\pi}{\omega\sqrt{\varepsilon}\rho}}$ and similarly for $H_0^{(+)}(\omega\sqrt{\varepsilon}\bar{\rho})$. In these formula we introduced the Fresnel coefficients $r(\omega,k_x):=r_\varphi,t(\omega,k_x):=t_\varphi$ associated with the wavevectors $k_x=\omega\sqrt{\varepsilon}\sin{\varphi},k_z=\omega\sqrt{\varepsilon}\cos{\varphi}$ and corresponding to a radiation at the angle $\varphi\in[-\pi/2,+\pi/2]$ (similar definitions are introduced for $r_{\bar{\varphi}},t_{\bar{\varphi}}$).\\
\indent In the region outside the cavity (for $z>D$), the SDP contribution corresponds well to the radiative field (far-field). This term completely overwhelms the singular term $\Phi_P$ which decays exponentially.\\
\indent The radiation field can be used to reconstruct an image of the field in the cavity using a microscope (with a oil immersion objective for matching the optical index. More precisely, according to the classical theory of optical imaging by a microscope \cite{Berthel}, the field $\Phi_{imag}(x')$ imaged conjugate with the plane $z=0$ is given by an integral of the type
\begin{eqnarray}
\Phi_{imag}(x')\simeq A \int_{-\omega NA}^{+\omega NA}dk_xt(\omega,k_x)\tilde{\Psi}_{c}(k_x)e^{-ik_xx'/M}
\end{eqnarray}
where, for simplicity, we neglect $\varepsilon"$ (i.e., we have $\sqrt{\varepsilon}=n$ where $n$ is the optical index inside and outside the cavity until the oil immersion objective) and where we have introduced the numerical aperture $NA=\sqrt{\varepsilon}\sin{\phi_{max}}$ associated with the objective lens, the magnification $M$ of the microscope, and where we have neglected the effects of optical aberrations (which do not play a major role here in the paraxial regime).  The constant $A$ is here irrelevant and depends on the various transmission coefficients in the microscope. On the other hand, using $\tilde{\Psi}_{c}(k_x)=\frac{i\omega I_0d}{2\pi }\frac{1}{1+re^{2ik_zD}}$ and expanding the term $1+re^{2ik_zD}$ in the vicinity of the poles $\pm k_{xP}$, we obtain 
\begin{eqnarray}
\Phi_{imag}(x')\simeq -A\frac{\omega I_0d t_P}{\frac{\partial(re^{2ik_zD})}{\partial k_x}|_{k_{xP}}}
\cdot\int_{-\omega NA}^{+\omega NA}dk_x\frac{i}{2\pi}[\frac{1}{k_{xP}-k_x}+\frac{1}{k_{xP}+k_x}]e^{-ik_xx'/M}
\end{eqnarray}
which can also be rewritten as
\begin{eqnarray}
\Phi_{imag}(x')\simeq A\int_{-\infty}^{+\infty}dx\Psi_P(x)\textrm{PSF}(x+x'/M)
\end{eqnarray}
 where we clearly see the singular field 
 \begin{eqnarray}
\Psi_P(x)=\frac{-\omega I_0dt_P}{\frac{\partial(re^{2ik_zD})}{\partial k_x}|_{k_{xP}}}e^{ik_{xP}|x|}
\end{eqnarray} (compare with Eq.~\ref{pole}) and the convolution function with the point spread function $\textrm{PSF}(u)=\frac{\sin{[\omega  NA u]}}{\pi u}$ associated with the diffraction attributable to the microscope.\\
\indent Clearly, what this shows is that the image field is a fingerprint of the singular field associated with the eigenmode propagating in the cavity. Thus, although the singular term is not directly observed, its signature remains visible in the far-field, which fully explains the observation of such waves in the image plane of the microscope in agreement with the results of Sharoglazova et al. \cite{Klaers2025}.

%%%%%%%%%%%%%%ù
\bibliographystyle{eplbib} 

\end{document}